\documentclass[sigconf, nonacm]{acmart}

\usepackage{hyperref}
\usepackage[linesnumbered,ruled,vlined]{algorithm2e}
\usepackage{caption}
\usepackage{subcaption}
\usepackage{float}
\usepackage{alltt}
\usepackage{multirow}
\usepackage{cleveref}
\usepackage{enumitem}
\usepackage{xspace}
\usepackage{tikz}

\newcommand*\circled[1]{\tikz[baseline=(char.base)]{
            \node[shape=circle,draw,inner sep=0.3pt] (char) {#1};}}
            
\usetikzlibrary{decorations.pathreplacing,calc}
\newcommand{\tikzmark}[1]{\tikz[overlay,remember picture] \node (#1) {};}

\newcommand{\stitle}[1]{\vspace{2pt}\noindent\textbf{#1}}

\theoremstyle{plain} 

\newtheorem{definition}{Definition}

\newcommand{\boexplain}{\textsf{BOExplain}\xspace}

\newcommand{\sqlml}{inference\xspace}

\newcommand{\bl}[1]{{{\textcolor{green}{\{\{BL: \bf #1\}\}}}\xspace}}
\newcommand{\jn}[1]{{{\textcolor{red}{\{\{JN: \bf #1\}\}}}\xspace}}
\newcommand{\ewu}[1]{{{\textcolor{red}{\{\{wu: \bf #1\}\}}}\xspace}}

\begin{document}

\title{Explaining Inference Queries with Bayesian Optimization}

\affiliation{
{\Large Brandon Lockhart\large{$^{\diamondsuit}$}  \qquad  {\Large Jinglin Peng\large{$^{\diamondsuit}$}}  \qquad {\Large Weiyuan Wu\large{$^{\diamondsuit}$}} \qquad {\Large Jiannan Wang\large{$^{\diamondsuit}$}} \qquad {\Large Eugene Wu\large{$^{\dag}$}}    } 
\and 
{\normalsize Simon Fraser University\Large{$^{\diamondsuit}$}} \,\,\, {\normalsize Columbia University\Large{$^{\dag}$}}  \\
{\small \{brandon\_lockhart, jinglin\_peng, youngw, jnwang\}@sfu.ca}  \,\,\, {\small ewu@cs.columbia.edu}
}

\begin{abstract}
Obtaining an explanation for an SQL query result can enrich the analysis experience, reveal data errors, and provide deeper insight into the data. Inference query explanation seeks to explain unexpected aggregate query results on inference data; such queries are challenging
to explain because an explanation may need to be derived from the source, training, or inference data in an ML pipeline.  In this paper, we model an objective function as a black-box function and propose \boexplain, a novel framework for explaining \sqlml queries using Bayesian optimization (BO). An explanation is a predicate defining the input tuples that should be removed so that the query result of interest is significantly affected. BO --- a technique for finding the global optimum of a black-box function --- is used to find the best predicate. We develop two new techniques (individual contribution encoding and warm start) to handle categorical variables.   We perform experiments showing that the predicates found by \boexplain have a higher degree of explanation compared to those found by the state-of-the-art query explanation engines. We also show that \boexplain is effective at deriving explanations for \sqlml queries from source and training data on a variety of  real-world datasets. \boexplain is open-sourced as a Python package at \url{https://github.com/sfu-db/BOExplain}
\end{abstract}

\pagestyle{plain} 

\maketitle

\section{Introduction}\label{sec:intro}


Data scientists often need to execute aggregate SQL queries on \emph{inference data} to inspect a machine learning (ML) model’s performance. We call such queries \emph{inference queries}, which can be seen as an SQL query whose expressions may perform model inference. Consider an inference dataset with four variables \textsf{(customer\_id, age, sex, M.predict(I))}, where \textsf{M.predict(I)} represents a variable where each value denotes whether the model $M$ predicts the customer will be a repeat buyer or not. Running the following \sqlml query will return the number of female (predicted) repeat  buyers:
\begin{alltt}
SELECT COUNT(*) FROM InferenceData as I
WHERE sex = 'female' and M.predict(I) = 'repeat buyer'
\end{alltt}

If the query result is surprising, e.g., the number of repeat buyers is higher than expected, the data scientist may seek an explanation. One popular explanation method is to find a subset of the input data such that when this subset is removed, and the query is re-executed, the unexpected result no longer manifests~\cite{wu13, roy14}. This method is known as a \textit{provenance} or \textit{intervention}-based explanation \cite{miao2019going}.

\begin{table}[t]\vspace{1em}
    \centering \small
    \begin{tabular}{|c||c|c|c|}
     \hline
    \multirow{2}{*}{}     & {\bf SQL Explain} & \multicolumn{2}{c|}{{\bf Inference Query Explain}} \\ \cline{3-4}
         &  {\bf \cite{wu13, roy14, roy15, abuzaid2020diff}} & {\bf Rain~\cite{wu2020complaint}} & {\bf BOExplain} \\ \hline \hline
    {\bf Inference Data} & Supported & Supported & Supported \\ \hline
    {\bf Training Data} & Not Supported  & Supported  & Supported \\ \hline
    {\bf Source Data} & Not Supported  & Not Supported  & Supported \\ \hline
    {\bf Explanation Type} & Coarse-grained & Fine-grained  &  Coarse-grained \\  \hline \hline
    {\bf Methodology} & White-box & White-box & Black-box \\ \hline
    \end{tabular}
    \caption{Comparison of \boexplain and existing approaches.}
    \label{tab:compare}
\end{table}


Specifically, there are two types of explanations in the intervention-based setting: fine-grained (a set of tuples) and coarse-grained (a predicate) \cite{melioututorial}. In this paper, we focus on  coarse-grained explanation. Predicates, unlike sets of tuples, provide a comprehensible explanation and identify common properties of the input tuples that cause the unexpected result. For the above example, it may return a predicate like  $\texttt{sex} = \texttt{`female'} \texttt{ AND } 20 \le \texttt{age} \le 25$ which suggests that if removing the young female customers from the inference data, the query result would look normal. Then, the data scientist can look into these customers more closely and conduct further investigation.

Generating an explanation from inference data can certainly help to understand the answer to an \sqlml query. However, an ML pipeline does not only contain inference data but also training and source data. The following example illustrates a scenario where an explanation should be generated from source data. 


\begin{figure*}[!ht]
    \centering
    \includegraphics[width=\textwidth]{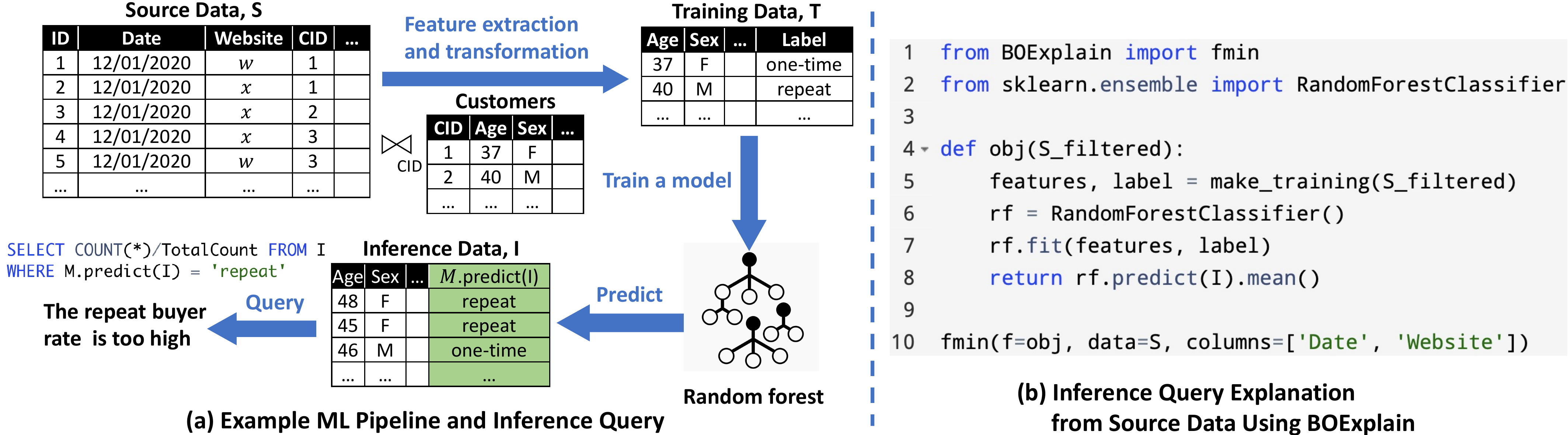}
    \caption{An illustration of using BOExplain to generate an explanation from source data in an ML pipeline.} 
    \label{fig:workflow}
\end{figure*}

\begin{example}\label{exa:explain-source}\it
CompanyX creates an ML pipeline (\Cref{fig:workflow}(a)) to predict repeat customers for a promotional event. CompanyX receives transaction records from several websites that sell their products and aggregates them into a \textit{source} data table $S$. Next, the user defined function (UDF) \textsf{make\_training($\cdot$)} extracts and transforms features into the \textit{training} dataset $T$. Finally, a random forest model is fit to the training data, and the model is applied to the inference dataset $I$ which updates it with a prediction variable, $M.predict(I)$. 

For validation purposes, the data scientist writes a query to compute the percentage of repeat buyers. The rate is  higher than expected, but she wants to double check that the result is not due to a data error.   In fact, it turns out that the source data $S$ contains errors during $\texttt{Date} \in [t_1, t_2]$, when the website $w$ had network issues; customers confirmed their transactions multiple times, which led to duplicate records in $S$.  The training data extraction UDF was coded to label customers with multiple purchases as repeat buyers, and labelled {\it all} of the $w$ customers during the network issue as repeats.  The model erroneously predicts every website $w$ customer as a repeat buyer, and thus leads to the high query result.  Ideally, the data scientist could ask whether the source data contains an error, and an explanation system would generate a predicate ($t_1 \le \texttt{Date} \le t_2 \texttt{ AND } \texttt{Website}=w$).
\end{example}


Unfortunately, existing SQL explanation approaches~\cite{wu13, roy14, roy15, abuzaid2020diff} are ill-equipped to address this setting (\Cref{tab:compare}) because they are based on analysis of the query provenance.  Although they can generate a predicate explanation over the inference data, the provenance analysis does not extend across model training nor UDFs, which are prevalent in data science workflows.  The recent system  Rain~\cite{wu2020complaint} generates fine-grained explanations for \sqlml queries.  It relaxes the \sqlml query into a differentiable function over the model's prediction probabilities, and leverages influence analysis~\cite{koh2017understanding} to estimate the query result's sensitivity to a training record. However, Rain returns training records rather than predicates, and estimating the model prediction sensitivity to group-wise changes to the training data remains an open problem.  Further, Rain does not currently support UDFs and uses a white-box approach that is less amenable to data science programs (\Cref{fig:workflow}(b)) that heavily incorporate UDFs.

As a first approach towards addressing the above limitations, and to diverge from existing white-box explanation approaches~\cite{wu13, roy14, roy15, abuzaid2020diff, wu2020complaint}, this paper explores a black-box approach towards \sqlml query explanation.   
\boexplain models \sqlml query explanation as a hyperparameter tuning problem and adopts Bayesian Optimization (BO) to solve it. In ML, hyperparameters (e.g., the number of trees, learning rate) control the training process and are tuned in an ``outer-loop'' that surrounds the model training process.  Hyperparameter tuning seeks to find the best hyperparameters that maximizes some model quality measure (e.g., validation score). \boexplain treats predicate constraints (e.g., $t_1, t_2, w$ in Example~\ref{exa:explain-source}) as hyperparameters, and the goal is to assign the optimal value to each constraint. By defining a metric that evaluates a candidate explanation's quality,  (e.g., the decrease of the repeat buyer rate), \boexplain finds the constraint values that correspond to the highest quality predicate.


A black-box approach offers many advantages for \sqlml query explanation.
In terms of \textbf{usability}, a data scientist can derive a predicate from any data involved in an ML pipeline rather than inference data only. Furthermore, its concise API design is similar to popular hyperparameter tuning libraries, such as scikit-optimize \cite{tim_head_2018_1207017} and Hyperopt \cite{bergstra2013hyperopt}, that many data scientists are already very familiar with. Figure~\ref{fig:workflow}(b) shows an example using \boexplain's API to solve \Cref{exa:explain-source}.  The data scientist wraps the portion of the program in an objective function \textsf{obj} whose input is the dataset to generate predicates for, and whose output is the repeat buyer rate that should be minimized.  She also provides hints to focus on the \textsf{Date} and \textsf{Website} variables. See Section~\ref{framework} for more details. 

In terms of \textbf{adaptability}, a black-box approach can
potentially be used to generate explanations for any data science workflow beyond \sqlml queries.  The current ML and analytics ecosystem is rapidly evolving.  In contrast to white-box approaches, which must be carefully designed for specific programs, \boexplain can more readily evolve with API, library, model, and ecosystem changes.

In terms of \textbf{effectiveness}, \boexplain builds on the considerable advances in BO by the ML community~\cite{shahriari2015taking}, to quickly generate high quality explanations. A secondary benefit is that BO is a progressive optimization algorithm, which lets \boexplain quickly propose an initial explanation, and improve it over time.

The key technical challenge is that existing BO approaches~\cite{bergstra2011algorithms,hutter2011sequential,snoek2013bayesian} cannot be naively adapted to explanation generation. In the hyperparameter tuning setting, categorical variables typically have very low cardinality (e.g., with 2-3 distinct values~\cite{nguyen2020bayesian}). In the query explanation setting, however, a categorical variable can have many more distinct values. To address this, we propose a categorical encoding method to map a categorical variable into a numerical variable.  This lets \boexplain estimate the quality of the categorical values that have not been evaluated.  We further propose a warm start approach so that \boexplain can prioritize predicates with more promising categorical values.




\smallskip\noindent In summary, this paper makes the following contributions: 
\begin{itemize}[leftmargin=*]
    \item We are the first to generate coarse-grained explanations from the training and source data to an \sqlml query. We argue for a black-box approach to \sqlml query explanation and discuss its advantages over a white-box approach.
    \item We propose \boexplain, a novel query explanation framework that derives explanations for \sqlml queries using BO. We develop two techniques (categorical encoding and warm start) to improve \boexplain's performance on categorical variables.  
    \item We show that  \boexplain can generate comparable or higher quality explanations than state-of-the-art SQL explanation engines (Scorpion~\cite{wu13} and MacroBase~\cite{abuzaid2020diff}) on SQL-only queries.  
    We evaluate \boexplain using \sqlml queries on real-world datasets showing that \boexplain can generate explanations for different input datasets with a higher degree of explanation than random search.
\end{itemize}

\section{Problem Definition}

\label{sec:prob_def}

In this section, we first define the SQL explanation problem, and subsequently describe the extension to inference query explanation.

\subsection{Background: SQL Explanation}


\stitle{Query.} We first define the supported queries. In this work, we focus on aggregation queries over a single table (the extension to multiple tables has been formalized in \cite{roy14}). An \textit{explainable query} is an arithmetic expression over a collection of SQL query results, as formally defined in Definition~\ref{def:query}.

\begin{definition}[Supported Queries]
\label{def:query}
Given a relation $R$, an \textit{explainable query} $Q=E(q_1,\dots,q_k)$ is an arithmetic expression $E$ over queries $q_1, \dots, q_k$ of the form
\begin{align*}
\small
q_i= & \textbf{\textsf{ SELECT}} \text{ agg}(\dots) \textbf{\textsf{ FROM }} R \\
 & \textbf{\textsf{ WHERE }} C_1 \textbf{\textsf{ AND/OR }}\dots\textbf{\textsf{ AND/OR }} C_m
\end{align*}
where agg is an aggregation operation and $C_j$ is a filter condition.
\end{definition}

\begin{example}\label{ex:query_ex}\it
Returning to the running example in Section~\ref{sec:intro}, the user queries the predicted repeat buyer rate. This can be expressed as $Q=q_1/q_2$, an arithmetic expression over $q_1$ and $q_2$ where
\begin{align*}\small
    q_1 &= \text{SELECT COUNT(*) FROM I WHERE M.predict(I)=`repeat buyer';} \\
    q_2 &= \text{SELECT COUNT(*) FROM I;}
\end{align*}
\end{example}

\stitle{Complaint.} After the user executes a query, she may find that the result is unexpected and \textit{complain} about its value. In this work, the user can complain about the result being too high or too low, as done in ~\cite{roy14}. We use the notation $dir=low$ ($dir=high)$ to indicate that $Q$ is unexpectedly high (low).

\begin{example}
In our running example, the user found the repeat buyer rate too high. Thus along with the query $Q$ from Example~\ref{ex:query_ex}, the user specifies $dir=low$ to indicate that $Q$ should be lower.
\end{example}

\stitle{Explanation.} After the user complains about a query result, \boexplain will return an explanation for the complaint. In this work, we define an explanation as a predicate over given variables. 


\begin{definition}[Explanation]
\label{def:explanation}
Given numerical variables $N_1, \dots, N_n$ and categorical variables $C_1, \dots, C_m$, an explanation is a predicate $p$ of the form $$p=l_1\le N_1\le u_1\wedge\dots\wedge l_n\le N_n\le u_n\wedge C_1=c_1\wedge\dots\wedge C_m=c_m.$$ The set of all such predicates forms the \textit{predicate space} $S$.
\end{definition}

\begin{example} \it The source data in Figure~\ref{fig:workflow} contains the variables Date and Website. An example explanation over these variables is $$12/01/2020 \le \text{Date} \le 12/10/2020 \wedge \text{Website}=w.$$
\end{example}

\stitle{Objective Function.} Next we define our objective function. The goal of our system is to find the best explanation for the user's complaint. Hence, we need to measure the quality of an explanation. For a predicate $p$, let $\sigma_{\neg p}(R)$ represent $R$ filtered to contain all tuples that do not satisfy $p$. We apply the query to $\sigma_{\neg p}(R)$ and get the new query result. If the user specifies $dir=low$, then the smaller the new query result is, the better the explanation is. Hence, we use the new query result as a measure of explanation quality. The objective function is formally defined in Definition~\ref{def:obj}.

\begin{definition}[Objective Function] 
\label{def:obj}
Given a predicate $p$, relation $R$, and query $Q=E(q_1,\dots,q_k)$, the objective function $obj(p, R, Q) \rightarrow \mathbb{R}$ applies $Q$ on the relation $\sigma_{\neg p}(R)$.
\end{definition}

With the definition of objective function, the problem of searching for the best explanation is equivalent to finding a predicate that minimizes or maximizes the objective function. 

\begin{definition}[SQL Explanation Problem]
\label{def:problem}
Given a relation $R$, query $Q=E(q_1,\dots,q_k)$, direction $dir$, and predicate space $S$, find the predicate \emph{$p^*=\operatorname*{arg\,min}_{p\in S}obj(p, R, Q)$}
if $dir=low$ (use $\operatorname*{arg\,max}$ if $dir=high$).
\end{definition}


It may appear that minimizing the above objective function runs the risk of overfitting to the user's complaint (perhaps with an overly complex predicate).  However, a regularization term can be placed within the objective function---for instance, SQL explanation typically regularizes using the number of tuples that satisfy the predicate~\cite{wu13}.  Since $Q$ is an arithmetic expression over multiple queries, one of those queries may simply be the regularization term. 

\subsection{Extension to Inference Query Explanation}



For inference query explanation, we focus on three input datasets that the user can generate explanations from: source, training, and inference\footnote{In general, any intermediate dataset is acceptable, however we focus on these three due to their prevalence and to simplify the paper.}. The query processing pipeline is as follows (Figure~\ref{fig:workflow}(a)):
\begin{enumerate}[leftmargin=*]
    \item Transform and featurize the source data into the training data.
    \item Train an ML model over the training data.
    \item Use the model to predict a variable from the inference dataset.
    \item Issue a query over the inference dataset.
\end{enumerate}

\smallskip\noindent From the above workflow, we can find that there are two differences between SQL and inference query explanations: 1) the query for inference query explanation is evaluated on the inference data with model predictions, and 2) in inference query explanation, the user may want an explanation for the input dataset at any step of the workflow (e.g., the source, training, or inference dataset), while SQL explanation only consider the query's direct input. 

We next formally define the scope of the errors that we seek to explain in Definition 5.

\begin{definition}[Scope of Errors]
This paper focuses on errors in the form of systematically mislabelled tuples that can be described using a predicate as defined in Definition~\ref{def:explanation}.
\end{definition}

We next extend the objective function from SQL explanation to inference query explanation. Let $Q$ be the query issued by the user over the updated inference data, with the same form as in Definition~\ref{def:query}. Let $R$ be the data that we want to derive an explanation from (it can be source, training, or inference data) and $p$ be an explanation (i.e., predicate) over $R$. We measure the quality of $p$ like in SQL explanation: filter the data by $p$, then get the new query result. Note that for inference query explanation, the query is issued over the updated inference data. Hence, we define $\mathcal{P}$ as the subset of the ML pipeline that takes as input the dataset $R$ that we wish to generate an explanation from, and that outputs the updated inference data which is used as input to the SQL query. The extended objective function is defined in Definition~\ref{def:obj-sqlml}.

\begin{definition}[Objective Function]
\label{def:obj-sqlml}
Given a subset of an ML pipeline $\mathcal{P}$, a predicate $p$, relation $R$, and query $Q$, the objective function $obj(p, R, \mathcal{P}, Q) \rightarrow \mathbb{R}$ feeds $\sigma_{\neg p}(R)$ through $\mathcal{P}$, and then applies $Q$ on the inference data.
\end{definition}

Finally, we define the inference query explanation problem.

\begin{definition}[Inference Query Explanation Problem]
\label{def:problem-sql-ml}
Given a relation $R$, query $Q$, direction $dir$, pipeline $\mathcal{P}$, and predicate space $S$, find the predicate \emph{$p^*=\operatorname*{arg\,min}_{p\in S}obj(p, R, Q, \mathcal{P})$} if $dir=low$ (use $\operatorname*{arg\,max}$ if $dir=high$).
\end{definition}

We assume that an explanation in the form of Definition~\ref{def:explanation} that performs well under the objective function in Definition~\ref{def:problem-sql-ml} is meaningful to the user. Hence, if $dir=low$ ($high$), the predicate $p^*$ that minimizes (maximizes) the objective function is considered optimal.

\section{The \boexplain Framework}\label{sec:background}
This section introduces Bayesian optimization (BO) and presents the \boexplain framework. 

\subsection{Background}\label{tpe}
Black-box optimization aims to find the global minima (or maxima) of a black-box function $f$ over a search space~$\mathcal{X}$, $$x^*=\min_{x\in\mathcal{X}}f(x).$$
BO is a sequential model-based optimization strategy to solve the problem, where \emph{sequential} means that BO is an iterative algorithm and \emph{model-based} means that BO builds surrogate models to estimate the behavior of $f$. 

\stitle{Tree-structured Parzen Estimator (TPE)}. TPE~\cite{bergstra2011algorithms, bergstra2013making} is a popular BO algorithm. It first initializes by evaluating $f$ on random samples from the search space. Then, it iteratively selects $x$ from the search space using an \emph{acquisition function} and evaluates $f(x)$. Let $D = \{(x_1, f(x_1)), (x_2, f(x_2)), \cdots, (x_t, f(x_t))\}$ denote the set of samples evaluated in previous iterations. TPE chooses the next sample as follows:
\begin{enumerate}[leftmargin=*]
    \item Partition $D$ into sets $D^g$ and $D^b$, where $D^g$ consists of the set of 
    $\gamma$-percentile points with the lowest $f(x)$ values in $D$, and $D^b$ consists of the remaining points ($\gamma$ is a user-definable parameter). Since the goal is minimize $f(x)$, $D^g$ is called the \emph{good-point set} and $D^b$ is called the \emph{bad-point set}. Intuitively, good points lead to smaller objective values than bad points. 
    \item Use Parzen estimators (a.k.a kernel density estimators) to build a density model $g(x)$ and $b(x)$ over $D^g$ and $D^b$, respectively. Intuitively, given an unseen $x^*$ in the search space, the density models $g(x^*)$ and $b(x^*)$ can return the probability of $x^*$ being a good and bad point, respectively. Note that separate density models $g(x)$ and $b(x)$ are constructed for each dimension of $\mathcal{X}$.
    \item Construct an acquisition function $g(x)/b(x)$ and select $x$ with the maximum $g(x)/b(x)$ to evaluate in the next iteration. 
    Intuitively, TPE selects a point that is more likely to appear in the good-point set and less likely to appear in the bad-point set.  
\end{enumerate}

\begin{figure}[t]
    \centering
    \includegraphics[width=1\linewidth]{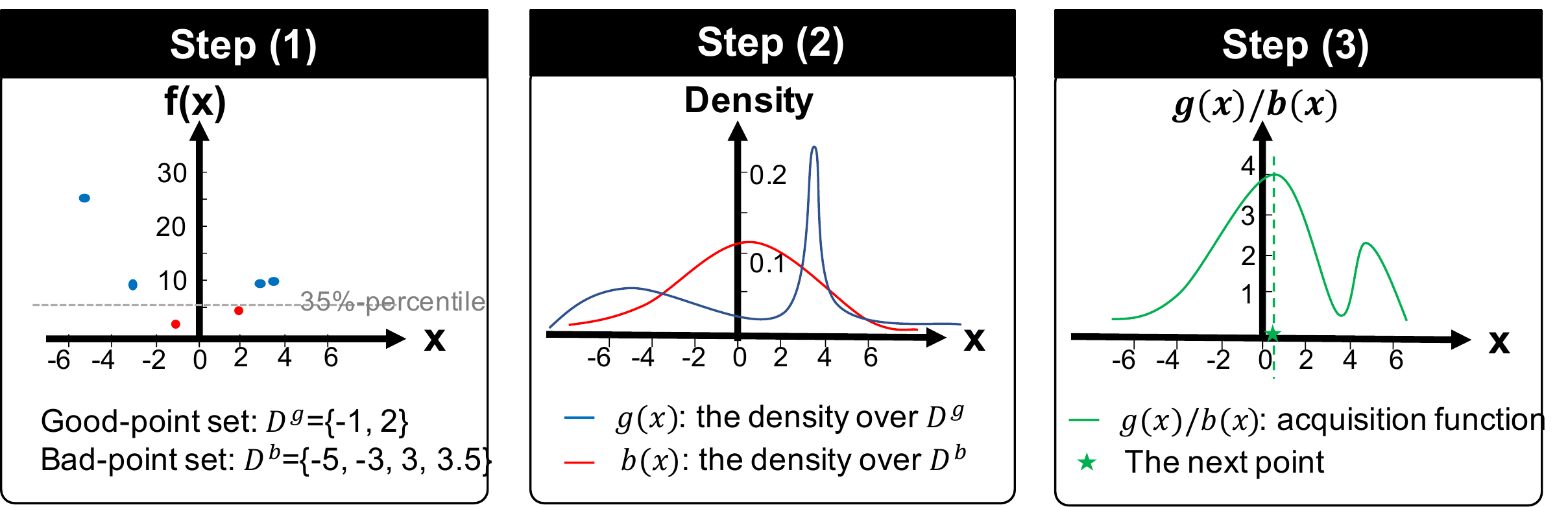}
    \caption{Suppose TPE has observed six points $D =\{(-5, 25), (-3, 9), (-1, 1), (2, 4), (3, 9), (3.5, 9.25)\}.$ This figure illustrates how TPE finds the next point to evaluate ($\gamma = 35\%$). }
    \label{fig:tpe-example}
\end{figure}

\begin{table}[t]
    \centering\small
\begin{tabular}{|c|c|c|c|c|c|}
    \hline
    {\bf Age} & {\bf Sex} & {\bf City} & {\bf State} & {\bf Occupation} & {\bf M.predict($I$)} \\
    \hline
    48 & F & Mesa & AZ & Athlete & repeat \\
    45 & F & Miami & FL & Artist & repeat \\
    46 & M & Mesa & AZ & Writer & one-time \\
    40 & M & Miami & FL & Athlete & repeat \\
    42 & F & Miami & FL & Athlete & repeat \\
    \hline
\end{tabular} \caption{An illustration of parameter creation.} \label{tab:inference}
\end{table}

\smallskip\noindent Figure~\ref{fig:tpe-example} illustrates the three steps. A complete introduction to TPE is given in Appendix~\ref{sec:appendix_tpe}.

\stitle{Categorical Variables.} TPE models categorical variables by using categorical distributions rather than kernel density estimation. Consider a categorical variable with four distinct values: \texttt{Website} $\in$ \{$w_1$, $w_2$, $w_3$, $w_4$\}. To build $g(\texttt{Website})$,
TPE estimates the probability of $w_i$ based on the fraction of its occurrences in $D^g$; the distribution is smoothed by adding $1$ to the count of occurrences for each value.  For instance, if the occurrences are 2, 0, 1, 0, then the distribution $g(\texttt{Website})$ will be $\{P(w_1), P(w_3), P(w_3), P(w_4)\} = \{3/7, 1/7, 2/7, 1/7\}$.

\subsection{Our Framework}\label{framework}

In this section, we describe the \boexplain framework. 




\stitle{Parameter Creation.} Given a predicate space, we need to map it to a 
parameter search space (the 
parameters and their domains). Suppose a predicate space is defined over variables $A_1, A_2, \cdots, A_n$. 

If $A_i$ is numerical (e.g., age, date), two parameters are created that serve as bounds on the range constraint. Specifically, the parameters $A_{i_{\text{min}}}$ and $A_{i_{\text{length}}}$ define the lower bound and the length of the range constraint, respectively. $A_{i_{\text{min}}}$ and $A_{i_{\text{length}}}$ have interval domains $[\min(A_i), \max(A_i)]$ and $[0, \max(A_i) - \min(A_i)]$, respectively.

If $A_i$ is categorical (e.g., sex, website), one categorical parameter is created with a domain consisting of all unique values in $A_i$.



\example\label{param_domain}{\it Suppose the user defines a predicate space over  \texttt{State} and \texttt{Age} in Table~\ref{tab:inference}. \boexplain creates three parameters: one categorical parameter for  \texttt{State} with domain \{AZ, FL\}, and two numerical parameters for \texttt{Age} with domains $[40, 48]$ and $[0, 8]$, respectively.
}

\begin{figure}
    \centering
    \includegraphics[width=0.48\textwidth]{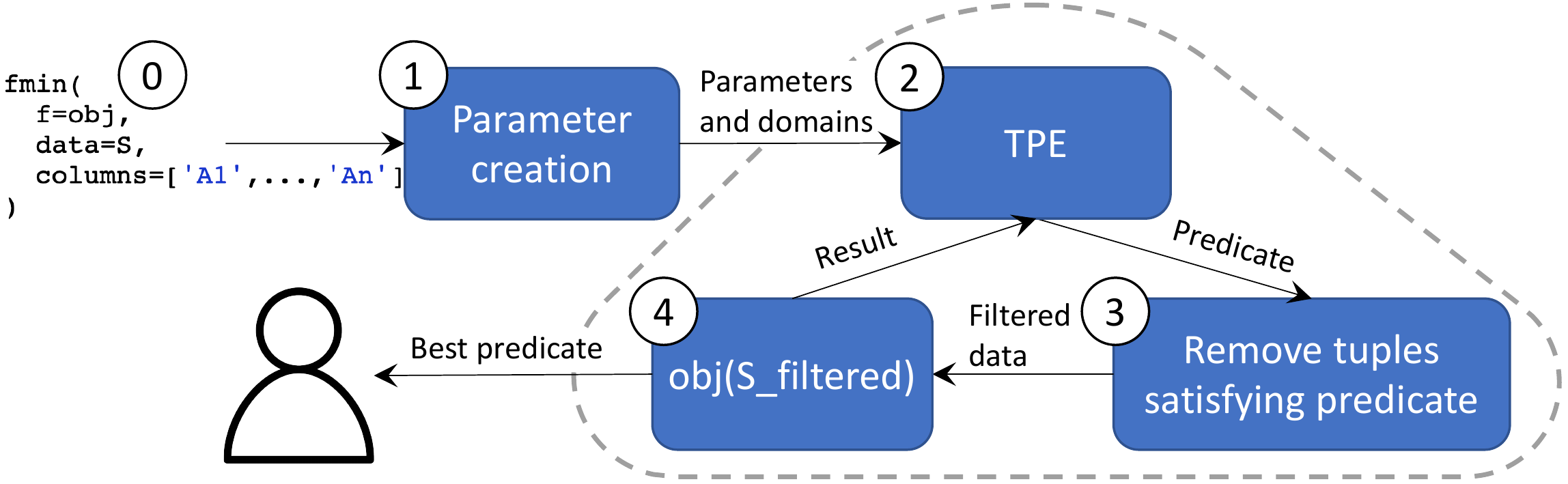}
    \caption{The \boexplain framework.}
    \label{fig:boexplain}
\end{figure}

\stitle{\boexplain Framework.}  \Cref{fig:boexplain} walks through the \boexplain framework.  In step \circled{0}, the user provides an objective  function $obj$, a relation $S$, and predicate variables $A_1, \dots, A_n$ (\Cref{fig:workflow}(b), line 10). Step \circled{1} creates the parameters and their domains. Step \circled{2} runs one iteration of TPE, starting with the parameters from step \circled{1}, and outputs a predicate.  Steps \circled{3} and \circled{4} evaluate the predicate by removing those tuples from the input dataset, and evaluating $obj$ on the filtered data.  The result is passed to TPE for the next iteration,  and possibly yielded to the user as an intermediate or final predicate explanation. 

Consider the example code in Figure~\ref{fig:workflow}(b). Once it is executed, \boexplain first creates three parameters: $\texttt{Date}_{{\text{min}}}$, $\texttt{Date}_{{\text{length}}}$, and $\texttt{Website}$ along with the corresponding domains. Then, it iteratively calls TPE to propose predicates (e.g., ``$12/01/2020 \leq \texttt{Date} \leq 12/02/2020$ \texttt{AND} $\texttt{Website} = w$''). \boexplain obtains S\_filtered by removing the tuples that satisfy this predicate from $S$. Next, it applies $\texttt{obj}(\cdot)$ to S\_filtered which will rerun the pipeline  (\Cref{fig:workflow}(a)) to compute the updated repeat buyer rate. The predicate and the updated rate are passed to TPE to use when selecting the predicate on the next iteration. This iterative process will repeat until the time budget is reached. When the user stops \boexplain, or when the optimization has converged, the predicate with the lowest repeat buyer rate is returned.

\stitle{Why Is TPE Suitable For Query Explanation?}  Recent work~\cite{bergstra2012random,li2020random,mania2018simple} has suggested that random search is a competitive strategy for hyperparameter tuning across a variety of challenging machine learning tasks.  However, we find that TPE is more effective for query explanation because it is designed for problems where similar parameter values tend to have similar objective values (e.g., model accuracy). TPE can leverage this property to prune poor regions of the search space.    As a trivial example, suppose a hyperparameter controls the number of trees in a random forest.  If values $10, 12, 14$ have resulted in a poor objective value, then TPE will down-weigh similar values (e.g., $9, 16$).

This property tends to hold in query explanation, because similar predicates tend to have similar objective values. For instance, we would expect that the predicate $\texttt{age} \in [10, 20]$ will exhibit a similar objective to $\texttt{age} \in [10, 19]$ and $\texttt{age} \in [10, 21]$; when the former has a poor objective value, the latter two may be pruned. When this property does not hold, \boexplain can still find the optimal predicate via the \textit{exploration} component of BO. BO balances two components for selecting a point to evaluate: 1) \textit{exploration} of the search space, and 2) \textit{exploitation} of points similar to previously well-performing points. \textit{Exploitation} may be ineffective if similar predicates do not perform similarly under the objective function, but \textit{exploration} will still test unpromising predicates, thus eventually leading BO to the optimal predicate.

\section{Supporting categorical variables} \label{sec:cat_enc}


In this section, we present our techniques to enable \boexplain to support categorical variables more effectively.

\subsection{Individual Contribution Encoding }\label{sec:indiv_cont_enc}

Recall that in Section~\ref{tpe}, TPE models numerical and categorical variables using kernel density estimation and categorical distribution, respectively. The advantage of kernel density estimation over a categorical distribution is that it can estimate the quality of unseen points based on the points that are close to them. To benefit from this advantage, we map a categorical variable to a numerical variable. We call this idea \emph{categorical encoding}. In the following, we  present our categorical encoding approach, called individual contribution (IC) encoding. 

A good encoding method should put \emph{similar} categorical values close to each other. Intuitively, two categorical values are similar if they have a similar contribution to the objective function value.  Based on this intuition, we rank the categorical values by their individual contribution to the objective function value. Specifically, consider a categorical variable $C$ with domain$(C)=\{c_1,\dots,c_n\}$. For each value $c_i$, we obtain the filtered dataset $\sigma_{C\ne c_i}(S)$ w.r.t. the predicate $C=c_i$. Next, the objective function is evaluated on the relation $\sigma_{C\ne c_i}(S)$ which returns a number. This number can be interpreted as the contribution of the categorical value on the objective function. After repeating for all values $c_i$, the categorical values are mapped to consecutive integers in order of their IC. \boexplain will then use a numerical rather than categorical variable to model $C$.

\example\label{ex:indiv_cont}{\it Suppose we would like an explanation from the inference data in Table~\ref{tab:inference}. Suppose the objective function value is the repeat buyer rate and the predicate space is defined over the Occupation variable. Note that the Occupation variable has the domain \{Athlete, Artist, Writer\}. The IC of Athlete is determined by removing the tuples where Occupation=``Athlete'' and computing the objective function on the filtered dataset, which gives $0.5$ (since only one of the two tuples in the filtered dataset is a repeat buyer). Similarly, the ICs of Artist and Writer are $0.75$ and $1$ respectively. Finally, we sort the categorical values by their objective function value and encode the values as integers: Athlete $\rightarrow1$, Artist $\rightarrow2$, Writer $\rightarrow3$.}



\subsection{Warm Start}\label{warm_start}

We next propose a warm-start approach to further enhance \boexplain's performance for categorical variables. 
Since an IC score has been computed for each categorical value, we can prioritize predicates that are composed of well performing individual categorical values. Rather than selecting $n_{\text{init}}$ points at random to initialize the TPE algorithm, we select the $n_{\text{init}}$ combinations of categorical values with the best combined score. More precisely, for a variable $C_i$, we consider the tuple pairs (variable value, IC) as computed in Section~\ref{sec:indiv_cont_enc}, $S_{IC}^i=\{(c_j, IC(c_j))\}_{j=1}^{n_i}$, where $n_i$ is the number of unique values in variable $C_i$. Next, we compute the d$-$ary Cartesian product and add the ICs for each combination $S_{IC}=S_{IC}^1\times\dots\times S_{IC}^d =\{((c_{i_1},\dots,c_{i_d}), IC(c_{i_1})+\dots+ IC(c_{i_d})) \mid i_j \in \{1,\dots,n_j\} \}$. 

\begin{example}\label{ex:ic}
The IC for values in the Occupation variable were computed in Example \ref{ex:indiv_cont}, $S_{ic}^{\text{Occupation}}=\{(\text{Athlete}, 0.5), (\text{Artist}, 0.75),\\ (\text{Writer}, 1)\}$, and for Sex we have $S_{ic}^{\text{Sex}}=\{(\text{F}, 0.5), (\text{M}, 1)\}$. Next we compute the combined IC score for each combination of predicates $S_{IC} = \{((\text{Athlete, F}), 1), \dots, ((\text{Writer, M}), 2)\}$. 
\end{example}

To see why adding ICs can be useful for prioritizing good predicates, suppose we want to minimize the objective function, and that $C_1=c_1$ and $C_2=c_2$ have small ICs. Then it is likely that $C_1=c_1 \wedge C_2=c_2$ has a small value. So we choose to sum the IC values as it encodes this property. Finally, we select $n_{\text{init}}$ valid predicates with the best combined IC score. Recall the user defines the direction that the objective function should be optimized. Therefore, we select the predicates with the smallest (largest) IC score if the objective function should be minimized (maximized). If the predicate also contains numerical variables, values are selected at random to initialize the range constraint parameters.

\begin{example} Continuing with Example~\ref{ex:ic}, recall that we want to minimize the objective function, so the smaller the combined IC score the better. Suppose $n_{\text{init}}=2$, then on the first and second iterations of BO, we evaluate the predicates $\texttt{Occupation}=\text{``Athlete''} \wedge \texttt{Sex}=\text{``F''}$ and $\texttt{Occupation}=\text{``Artist''} \wedge \texttt{Sex}=\text{``F''}$ respectively. Note that $\texttt{Occupation}=\text{``Athlete''} \wedge \texttt{Sex}=\text{``F''}$ is the best predicate, so adding IC scores can prioritize good explanations.
\end{example}

\subsection{Putting Everything Together}

We lastly present the full \boexplain algorithm in Algorithm~\ref{algo:boexplain}. First, the ICs for the categorical variables are computed in lines 1-3. Next, the parameters and domains are created in line 4. In line 5, the IC values are used to prioritize predicted high quality predicates, and in line 6 TPE is initialized for $n_{\text{init}}$ iterations with the predicted high quality predicates. Starting from line 7, we use a model to select the next points. In line 8, the previously evaluated points are split into good and bad groups based on $\gamma$. Next, from line 9, a value is selected for each parameter. In lines 10 and 11, distributions of the good and bad groups are modelled, respectively. In line 12, points are sampled from the good distribution, and the sampled point with the largest expected improvement is selected as the next parameter value (line 13). In line 15, the objective function is evaluated based on the parameter assignment, and the set of observation-value pairs is updated.


\begin{algorithm}[t]
\caption{\boexplain}
\label{algo:boexplain}
\DontPrintSemicolon
\SetKwInOut{Initialize}{Initialize}
\SetAlgoLined
\KwIn{Objective function $obj$, data $S$, variables $A_1,\dots,A_n$}
\KwOut{A predicate and the corresponding objective value}
\ForEach{categorical variable $C$}{
    Compute the IC of all unique values in $C$\;
}
\textbf{Create} the parameters and domains\;
Compute the predicted high quality combinations based on IC for the warm start\;
\textbf{Initialize TPE:} Perform $n_{\text{init}}$ iterations using a warm start to  create \tikzmark{right}  $D_{n_{\text{init}}} = \{(\textbf{x}_i , obj(\sigma_{\neg\textbf{x}_i}(S))\}_{i=1}^{n_{\text{init}}}$.\; 
\For{$t\leftarrow n_{\text{init}}$ \KwTo $n_{\text{iter}}$}{
    Split $D_t$ into $D_t^g$ and $D_t^b$ based on $\gamma$\;
    \For{$i\leftarrow 1$ \KwTo $d$}{
        Estimate $g(x)$ on the i$th$ dimension of $D_t^g$\;
        Estimate $b(x)$ on the i$th$ dimension of $D_t^b$\;
        Sample $n_{EI}$ points from $g(x)$\;
        Find the sample $x_{t+1}$ with the highest $g(x)/b(x)$\;
    }
    Update $D_{t+1} \leftarrow D_t\cup\{(\textbf{x}_{t + 1}, obj(\sigma_{\neg\textbf{x}_{t+1}}(S)))\}$ 
}
 \KwRet{$(\textbf{x}, obj(\sigma_{\neg\textbf{x}}(S)))\in D_{n_{\text{iter}}}$ with the best objective value} 
\end{algorithm}

\section{Experiments}\label{exp}

\begin{figure*}[!htbp]
    \centering
    \includegraphics[width=\textwidth]{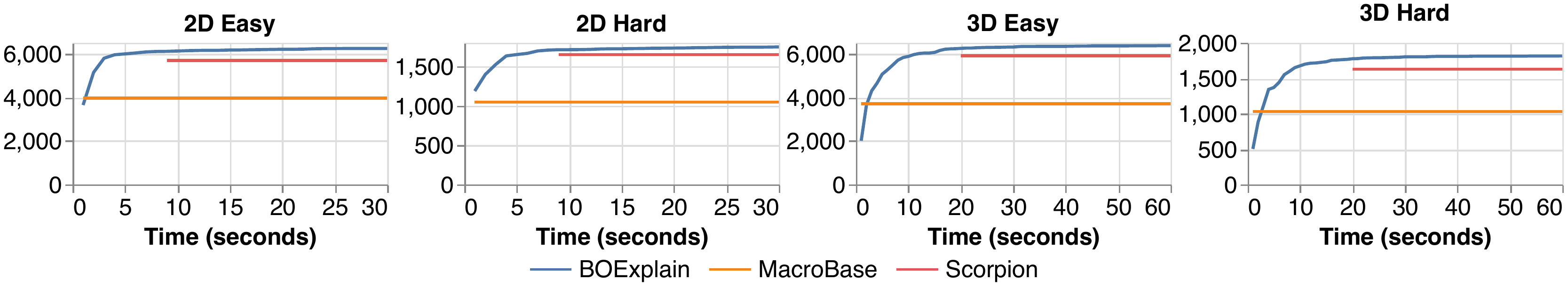}
    \caption{Performance comparison with Scorpion and MacroBase. The goal is to maximize the objective function.}
    \label{fig:scorpion}
\end{figure*}

Our experiments seek to answer the following questions. (1) How does \boexplain compare to current state-of-the-art query explanation engines for numerical variables? (2) Are the IC encoding and warm start heuristics effective? (3) How effective is \boexplain at deriving explanations from source and training data? (4) Can \boexplain generate useful explanations for real corrupted datasets?

\subsection{Experimental Settings}
\subsubsection{Baselines}
For SQL-only queries, we compare \boexplain with the explanation engines Scorpion \cite{wu13} and MacroBase \cite{abuzaid2020diff, bailis2017macrobase} which return predicates as explanations. For \sqlml queries, no predicate-based explanation engines exist, so we compare with a random search baseline~\cite{bergstra2012random} and Hyperband~\cite{li2017hyperband}. 

\stitle{Scorpion \cite{wu13}} is a framework for explaining group-by aggregate queries. The authors define a function to measure the quality of a predicate, which can be implemented as \boexplain's objective function. Each continuous variable’s domain is split into 15 equi-sized ranges as set in the original paper. We use the author's open-source code\footnote{\url{https://github.com/sirrice/scorpion}} to run the Scorpion experiments.

\stitle{MacroBase \cite{bailis2017macrobase} }  (later, the DIFF operator \cite{abuzaid2020diff}) is an explanation engine that considers combinations of variable-values pairs, similar to a \texttt{CUBE} query \cite{gray1997data}, as candidate explanations. 
In Section 2.3 of \cite{abuzaid2020diff}, the authors describe how to use the DIFF operator with Scorpion's objective function. We implemented it using the author's open-source code\footnote{\url{https://github.com/stanford-futuredata/macrobase}}. The user needs to discretize numerical variables; we tuned the bin size from 2 to 15 and report the best result.

In \cite{abuzaid2020diff}, MacroBase was shown to outperform other explanation engines including Data X-ray \cite{wang2015data} and Roy and Suciu \cite{roy15}, and so we do not compare with these approaches.

\stitle{Random} search is a competitive method for hyperparameter tuning \cite{bergstra2012random}. The parameters are chosen independently and uniformly at random from the domains described in Section~\ref{framework}.

\stitle{Hyperband}~\cite{li2017hyperband} is an exploration-based optimization strategy that speeds up random search through adaptive resource allocation and early-stopping.


\subsubsection{Real-world Datasets and ML Pipelines}
\begin{figure}[tb]
    \centering
    \includegraphics[width=0.45\textwidth]{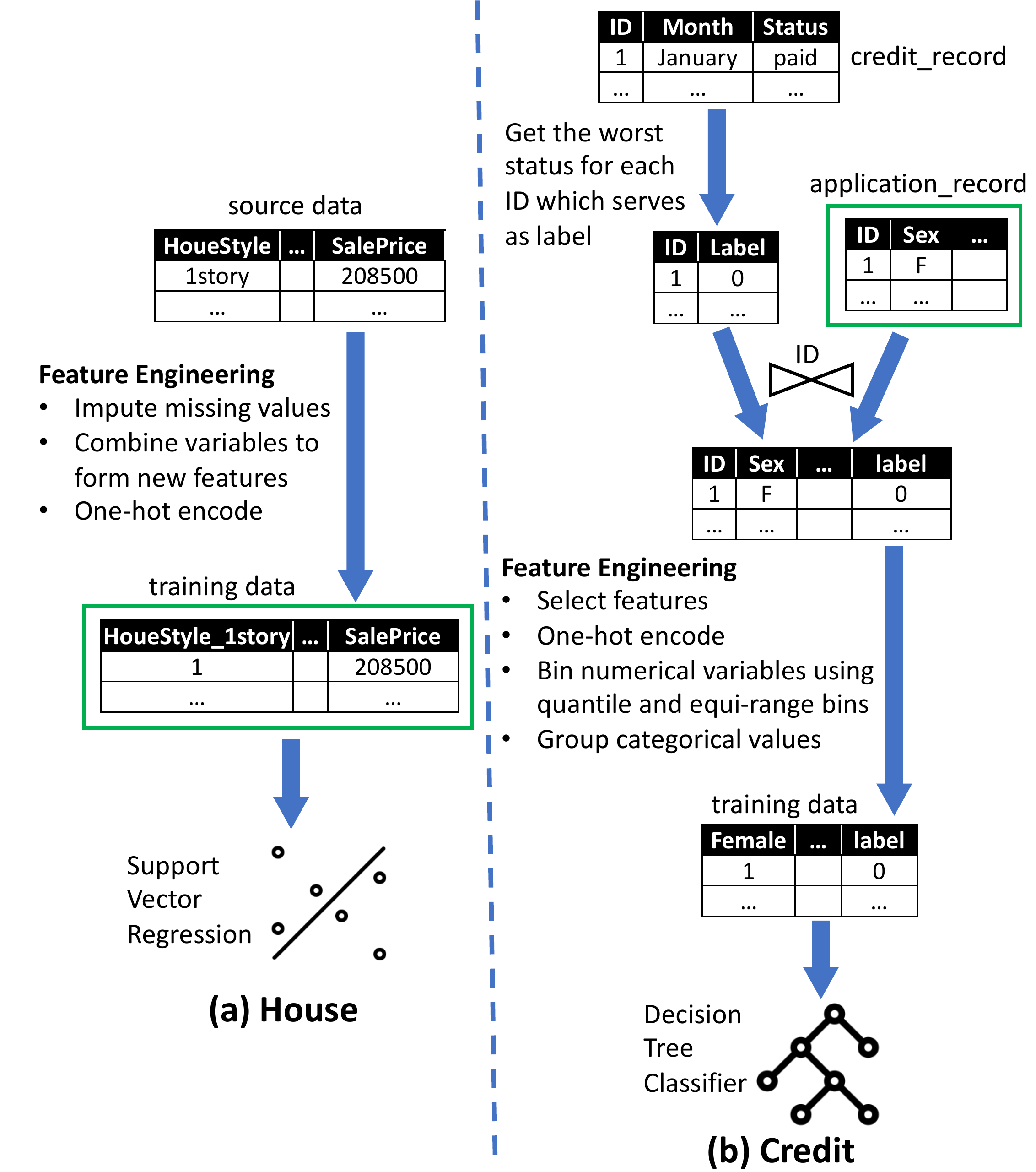}
    \caption{ML Pipelines for House and Credit. The green box indicates where an explanation is generated from.}
    \label{fig:pipeline}
\end{figure}

The following lists the five real-world datasets used in our experiments. We visualize the pipelines for House and Credit in Figure~\ref{fig:pipeline}, and put a green box around the data where an explanation is derived in each pipeline.
For House and Credit we inject synthetic errors. For Credit and Amazon, an explanation is derived from source data, for House, and German, an explanation is derived from training data, and for NYC, an explanation is derived from inference data.

\stitle{House} price prediction \cite{de2011ames}. This dataset was published already split into training (1460 rows) and inference (1459 rows) tables. It contains 79 variables of a house which are used to train a support vector regression model to predict the house price. The pipeline denoting how to prepare the data for modelling is given in Figure~\ref{fig:pipeline}(a).

\stitle{Credit} card approval prediction\footnote{\url{https://www.kaggle.com/rikdifos/credit-card-approval-prediction}}. The source data consists of two tables: \texttt{application\_record} (438,557 rows, 18 variables), which contains information about previous applicants, and \texttt{credit\_record} (1,048,575 rows, 3 variables), which stores the applicants' credit history. The pipeline to prepare the data for modelling is given in Figure~\ref{fig:pipeline}(b), and a 
 decision tree classifier is trained to predict whether a customer will default on their credit card payment. We set aside 20\% of the data to use for the \sqlml query, and 80\% for training.

\stitle{Amazon} product reviews~\cite{Ramirez2019}. This dataset contains 6928 reviews of Amazon products with ground truth and crowdsourced binary labels. 80\% of the reviews are used for training with labels formed from the majority vote of the crowdsourced labels, and 20\% for testing with the ground truth labels. We encode the reviews using Count Vectorization and train a support vector classifier.

\stitle{German} credit risk~\cite{Dua:2019}. This dataset contains 19 variables of 1000 bank customers with each customer labelled as having good or bad credit risk. We one-hot encode the categorical variables, do an 80-20 train-test split, and train an XGBoost classifier.

\stitle{NYC} yellow taxi dataset\footnote{\url{https://www1.nyc.gov/site/tlc/about/tlc-trip-record-data.page}}. This dataset contains taxi trip information for every yellow taxi trip in New York City. Following the setup in \cite{baier2020handling}, we predict the hourly demand by region for the 20 most frequent regions, and use the features weekday, region, demand of previous 24 hours, and cosine/sine features to encode that hours are cyclical. We train an XGBoost regressor on data from January and February, 2020, and perform inference on data from March 2020.

\subsubsection{Metrics}

To measure the quality of an explanation, we plot the best objective function value achieved by each time point $t$. 
For Scorpion and MacroBase we plot, the objective function value corresponding to their output predicate as a line that begins when the system finishes. To evaluate the effectiveness at identifying data errors, we measure the F-score, precision, and recall in the experiments on the House and Credit datasets. We synthetically corrupt data defined by a predicate, and use that data as ground truth. Precision is the number of selected corrupted tuples divided by the total number of selected tuples. Recall is the number of selected corrupted tuples divided by the total number of corrupted tuples. F-score is the harmonic mean of precision and recall. For \boexplain, Random, and Hyperband, each result is averaged over 10 runs.


\subsubsection{Implementation} \boexplain was implemented in Python 3.9. The code is open sourced at \url{https://github.com/sfu-db/BOExplain}. We modify the TPE algorithm in the Optuna library~\cite{akiba2019optuna} with our optimization for categorical variables. The ML models in Section~\ref{exp:real_world} are created with sklearn. The experiments were run single-threaded on a MacBook Air (OS Big Sur, 8GB RAM). In the TPE algorithm, we set $n_{init}=10$, $n_{ei}=24$, and $\gamma=0.1$ for all experiments.

\subsection{Explaining SQL-Only Queries}\label{exp:comparison}

\sloppy

To compare \boexplain, Scorpion, and MacroBase, we replicate the experiment from Section 8.3 of Scorpion's paper~\cite{wu13}, using the same datasets, query, and objective function. Note that MacroBase explicitly aims to optimize Scorpion's objective function, as described in Section 2.2 of ~\cite{abuzaid2020diff}.
The dataset consists of a single \textit{group by} variable $A_d$, an aggregate variable $A_v$, and search variables $A_1, \dots, A_n$ with $\text{domain}(A_i) = [0, 100]\subset \mathbb{R}$, $i\in[n]$. $A_d$ contains $10$ unique values (or $10$ groups) each corresponding to $2000$ tuples randomly distributed in the $n$ dimensions. $5$ groups are outlier groups and the other $5$ are holdout groups. Each $A_v$ value in a holdout group is drawn from $\mathcal{N}(10, 10)$. Outlier groups are created with two $n$ dimensional hyper-cubes over the $n$ variables, where one is nested inside the other. The inner cube contains $25\%$ of the tuples and $A_v\sim\mathcal{N}(\mu, 10)$, and the outer cube contains $25\%$ of the tuples in the group and $A_v\sim\mathcal{N}(\frac{\mu + 10}{2}, 10)$, else $A_v\sim\mathcal{N}(10, 10)$. $\mu$ is set to $80$ for the ``easy'' setting (the outliers are more pronounced), and $30$ for the ``hard'' setting (the outliers are less pronounced). The query is $\text{SELECT SUM}(A_v) \text{ FROM synthetic GROUP BY } A_d$.
The arithmetic expression over the SQL query is defined in Section $3$ of \cite{wu13} that forms an objective function to be maximized. The penalty $c=0.2$ was used to penalize the number of tuples removed as described in Section 7 of \cite{wu13}. We used $n=2$ and $n=3$ since $3$ is the maximum number of variables supported by MacroBase.

\fussy

The results are shown in Figure~\ref{fig:scorpion}. \boexplain outperforms Scorpion and MacroBase in terms of optimizing the objective function in each experiment. This is because \boexplain can refine the constraint values of the range predicate which enables it to outperform Scorpion and MacroBase which discretize the range. The results are the same in the easy and hard settings. MacroBase performs poorly because the predicates formed by discretizing the variable domains into equi-sized bins, and computing the cube, do not optimize this objective function. This exemplifies a known limitation of MacroBase that binning continuous variables is difficult \cite{abuzaid2020diff}.

\boexplain also outperforms Scorpion in terms of running time. \boexplain achieves Scorpion's objective function value in around half the time on each experiment.


\stitle{Note.} The focus of this paper is  \emph{not} on SQL-only queries, thus we did not conduct a comprehensive comparison with Scorpion and MacroBase. This experiment aims to show that a black-box approach (\boexplain) can even outperform white-box approaches (Scorpion and MacroBase) for SQL-only queries in some situations.


\subsection{Explaining Inference Queries}\label{exp:real_world}


In this section, we evaluate \boexplain's efficacy at explaining \sqlml queries from source and training data. In Section~\ref{exp:house}, we investigate \boexplain's approach for categorical variables on House, and in Section~\ref{exp:credit}, we evaluate \boexplain in a complex ML pipeline on Credit, where an explanation is derived from source data.

\begin{figure}[tb]
    \centering
    \includegraphics[width=0.5\textwidth]{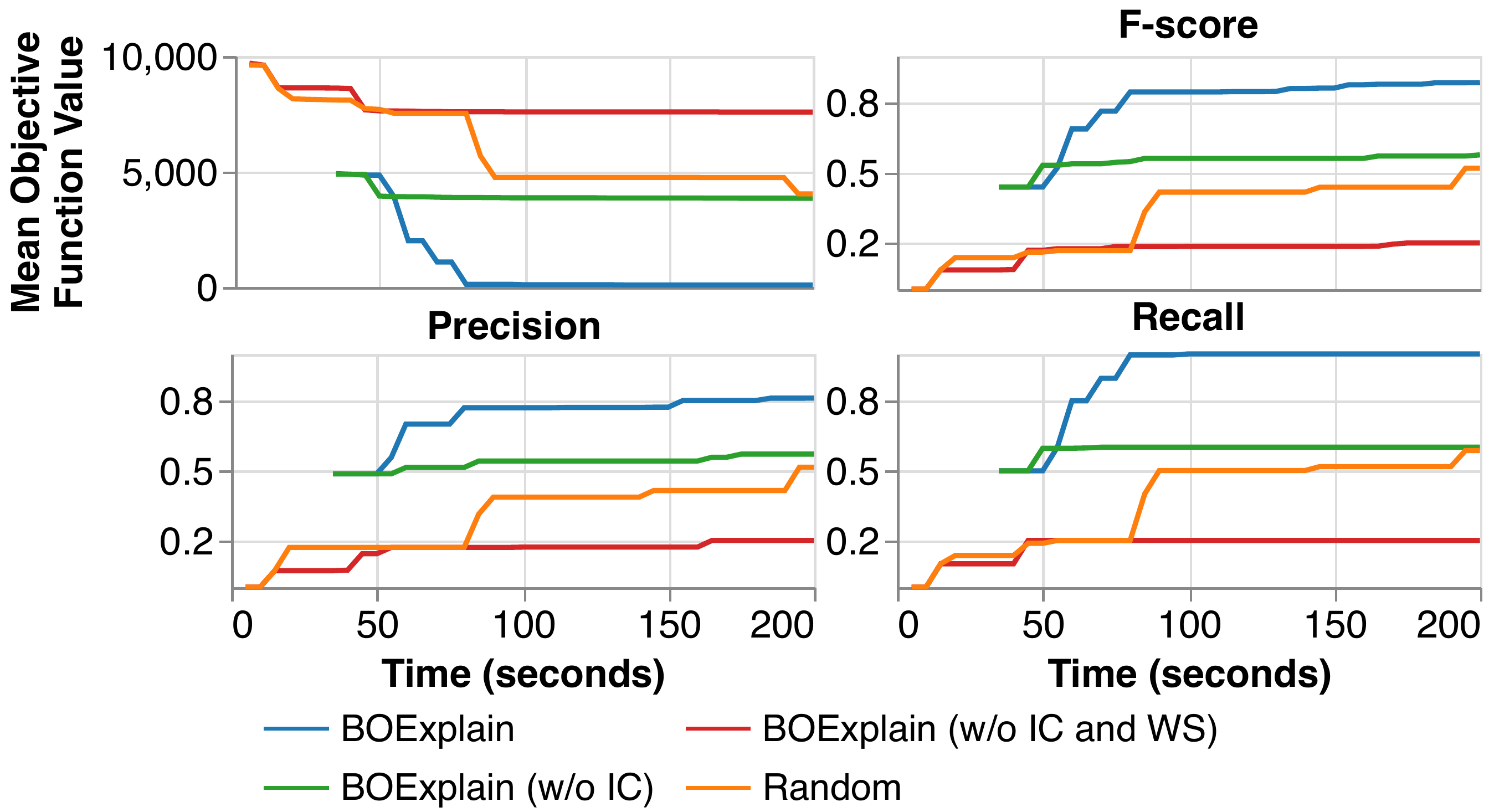}
    \caption{House: best objective function value, F-score, precision, and, and recall, found at each 5 second increment averaged over 10 runs. The goal is to minimize the objective function. (IC = Individual Contribution Encoding, WS = Warm Start)}
    \label{fig:house_results}
\end{figure}

\begin{figure*}[tb]
    \centering
    \includegraphics[width=\textwidth]{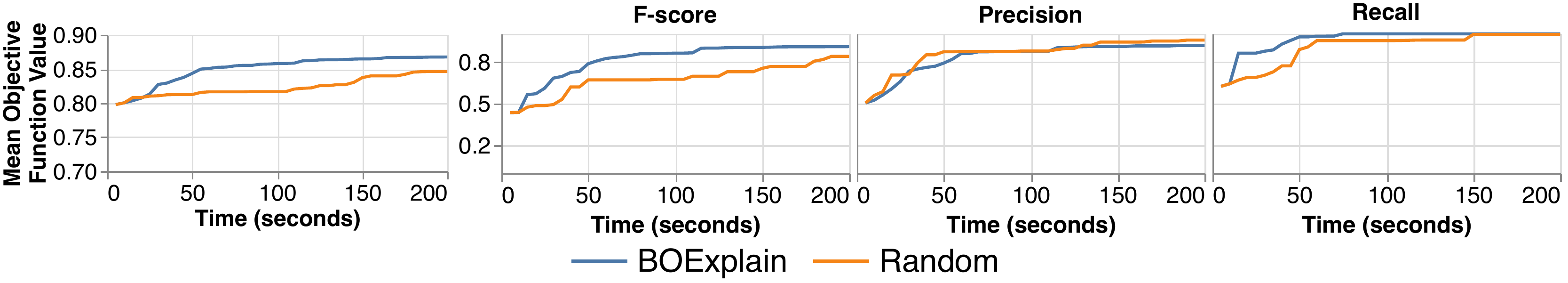}
    \caption{Credit: best objective function value, F-score, precision, and recall found at each 5 second increment, averaged over 10 runs. The goal is to maximize the objective function; larger values are better.}
    \label{fig:credit_results}
\end{figure*}



\subsubsection{\bf Supporting Categorical Variables} \label{exp:house}
\sloppy 
In this experiment, we assess \boexplain's method for handling categorical variables on House. The data is corrupted by setting the tuples satisfying \texttt{Neighbourhood}=\texttt{`CollgCr'} $\wedge $ \texttt{Exterior1st}=\texttt{`VinylSd'} $\wedge$ $2000 \le \texttt{YearBuilt} \le 2010$ to have their sale price multiplied by $10$, affecting 6.16\% of the data. We query the average predicted house price and seek an explanation for why it is high. To assess \boexplain's efficacy at removing the corrupted tuples, we define the objective function to minimize the distance between the queried result on the passed data and the result of the query issued on the data with the corrupted tuples removed. We use two categorical search variables Neighbourhood and Exterior1st which have 25 and 15 distinct values respectively, and one numerical search variable YearBuilt which has domain [1872, 2010]. The search space size is $7.25\times 10^6$.

\fussy


In this experiment, we compare three strategies for dealing with categorical variables. The first, \boexplain, is our algorithm with both of the IC encoding and warm-start (WS) optimizations proposed in Section~\ref{sec:cat_enc}. To determine whether encoding categorical values to integers based on IC and using a numerical distribution is effective, we consider a second approach, BOExplain (w/o IC), which uses the warm start optimization from Section~\ref{warm_start}, but uses the TPE categorical distribution to model the variables rather than encoding. The third, BOExplain (w/o IC and WS), is \boexplain without any optimizations. 

Each method is run for 200 seconds, and the results are shown in Figure \ref{fig:house_results}. The benefit of the warm start is apparent since \boexplain and \boexplain (w/o IC) outperform the other baselines much sooner. Also, \boexplain significantly outperforms \boexplain (w/o IC) which shows that encoding the categorical values, and using a numerical distribution to model the parameter, leads to BO learning the good region which can optimize the objective function when exploited. The F-score, precision, and recall also demonstrate how \boexplain can significantly outperform the baselines. In this experiment, \boexplain completed on average 274.3 iterations, whereas random completed 1148.4 iterations.

\stitle{Hyperband Experiment.} 
To evaluate our choice of using TPE, we also compare with a Hyperband implementation. For Hyperband, we use the data sample size as the resource for successive halving. To compare fairly with TPE, we run Hyperband for 200 seconds. We start with a random sample of 12.5\% of the data and randomly select predicates to evaluate by the objective function. Next, we select the 50\% best performing predicates, and evaluate their quality on a sample size of 25\%. This repeats until we evaluate the best predicates on 100\% of the data, and finally output the best predicate. The objective function value of the best found predicate averaged over 10 runs is $7666.03$, whereas for the TPE-based implementation it is $90.74$. Since the goal is to minimize the objective function, TPE performed better. The reason is that TPE with our proposed optimizations for categorical variables prioritized promising predicates early on in the search, whereas Hyperband's exploration-based search strategy could not find good quality predicates as quickly.





\subsubsection{\bf Explanation From Source Data}\label{exp:credit}


In the last experiment, we derive an explanation from source data on Credit. We corrupt the source data by setting all applicant records satisfying $-23000\le\texttt{DAYS\_BIRTH}\le-17000 \wedge 2\le\texttt{CNT\_FAM\_MEMBERS}\le3$ to have a ``bad'' credit status, which affects $20.1\%$ of the data. Corrupting the data decreases the accuracy of the model, and we define the objective function to increase the model accuracy. We derive an explanation from the source data table \texttt{application\_record} with the variables DAYS\_BIRTH and CNT\_FAM\_MEMBERS which have domains [-25201, -7489] and [1, 15], respectively, and the size of the search space is $7.06\times 10^{10}$.

The experiment is run for 200 seconds, and the results are shown in Figure~\ref{fig:credit_results}. On average, \boexplain completes 246.8 iterations and random search completes 319.6 iterations during the 200 seconds. \boexplain significantly outperforms Random at optimizing the objective function, as \boexplain on average attains an objective function value at 51 seconds that is higher than the average value Random attains at 200 seconds. This shows that exploiting promising regions can lead to better explanations, and that \boexplain is effective at deriving explanations from source data that passes through an ML pipeline. Although random search can find an explanation with high precision, \boexplain significantly outperforms Random in terms of F-score.

\subsection{Case Studies}\label{sec:case_studies}
To understand how \boexplain performs on real workloads, we present three case studies in this section. These case studies use  \boexplain to derive an explanation from three real-world datasets under realistic settings. The derived explanations are insightful, which show \boexplain's effectiveness in real-world applications.

\stitle{Crowdsourced Mislabels.} With the Amazon dataset, the inference accuracy is 93.75\%. To investigate whether mislabelled reviews decrease the accuracy, we define the objective function to increase the accuracy. We derive a predicate over the variables Country and TextWordCount (the number of words in the review) in the source dataset. After running \boexplain for 60 seconds, the output predicate is $\texttt{Country} = \text{"Turkey"} \wedge 101 \le \texttt{TextWordCount} \le 221$, which increases the test accuracy to 95\%. Upon further inspection, we found that the labelling accuracy of the training data is $93\%$, but the labelling accuracy of the tuples satisfying the returned predicate is $90\%$. Hence, \boexplain identified that labellers from Turkey were more likely to mislabel long reviews, which degraded the model.

\stitle{Bias.} With the German credit dataset, the predicted rate of good credit risk for individuals 25 years old or older is $76\%$, and for individuals under 25 years old it is $57\%$, hence this dataset is biased. We define the objective function to minimize the predicted rate of good credit risk between those over and under 25 years old, and derive a predicate from the training data with the search variables Purpose and DurationInMonths (duration of the loan). 

We ran \boexplain for 60 seconds, and the returned predicate is $\texttt{Purpose} = \text{"car (new)"} \wedge 20 \le \texttt{DurationInMonths} \le 50$. After removing the tuples satisfying this predicate, the predicted rate of having good credit risk is 74\% and 68\% for people over and under 25, respectively, which is significantly less biased. Moreover, in the training data, the overall good credit risk rate is 71\% and 60\% for people over and under 25, respectively, however for the tuples satisfying the predicates, the rates are 57\% and 17\%. Therefore, \boexplain identified that long-term loans for new cars are greatly biased in favour of people over 25 years old.

\stitle{Data Drift.}  With the NYC dataset, the mean-squared error (MSE) of taxi trip durations is 1801.03 on the training data and 10658.49 on the inference data. The objective function is to minimize the MSE on the inference data, and we search for an explanation over the inference data with the search variables Region and PickUpDateTime.

After running \boexplain for 30 seconds, the returned predicate is $\texttt{Region} = \text{"230"} \wedge \text{2020-03-12 21:00:00} \le \texttt{PickUpDatetime} \le \text{2020-03-31 23:00:00}$. When the tuples that satisfy this predicate are removed, the MSE on the inference data is 8706.13. In fact the MSE of the tuples satisfying the predicate is 18151.93, which is much higher than the overall MSE of the inference data (10658.49). Thus, \boexplain identified a region and time period that underwent significant drift as a result of the COVID-19 pandemic.
\vspace{-.5em}

\section{Related Work}\label{sec:related_work}

Our work is mainly related to query explanation, ML pipeline debugging, and Bayesian optimization.

\stitle{Query Explanation. }
\boexplain is most closely related to Scorpion~\cite{wu13} and the work of Roy and Suciu~\cite{roy14}. Both approaches define explanations as predicates. Scorpion uses a space partitioning and merging process to find the predicates, while Roy and Suciu~\cite{roy14} use a data cube approach. Both systems make assumptions about the aggregation query's structure in order to benefit from their white-box optimizations.  In contrast, \boexplain supports complex queries, model training, and user defined functions.  Further, \boexplain is a progressive algorithm that improves the explanation over time.   Variations of these ideas include the DIFF operator~\cite{abuzaid2020diff}, explanation-ready databases~\cite{roy15}, and counterbalances~\cite{miao2019going}. 
Finally, a number of specialized systems focus on explaining specific scenarios, such as streaming data~\cite{bailis2017macrobase}, map-reduce jobs~\cite{khoussainova2012perfxplain}, online transaction processing workloads \cite{yoon2016dbsherlock}, cloud services~\cite{roy2015perfaugur}, and range-radius queries~\cite{savva2018explaining}.

 Another related concept is the online analytical processing (OLAP) data cube \cite{gray1997data} which is used to explore and discover insights about subsets of multidimensional queries. Much previous work has been dedicated to providing the user with more meaningful and efficient exploration of the data cube \cite{sarawagi1998discovery, sathe2001intelligent, sarawagi2000i3}. Other work has used the \textit{data cube} concept to further understand the results of ML models \cite{chen2005prediction, chung2019slice, pastor2021looking}.
However, \boexplain is different from the cube-based approaches in two aspects. First, \boexplain can generate explanations from not only inference data but also training and source data. Second, \boexplain does not need to discretize numerical variables.

\stitle{ML Pipeline Debugging.}
Rain~\cite{wu2020complaint} is designed to resolve a user's complaint about the result of an \sqlml query by removing a set of tuples that highly influence the query result.
In contrast, \boexplain removes sets of tuples satisfying a predicate, which can be easier for a user to understand. In addition, \boexplain is more expressive, and supports UDFs, data science workflows, and pre-processing functions. Data X-Ray \cite{wang2015data} focuses on explaining systematic errors in a data generative process.
Other systems debug the configuration of a computational pipeline \cite{lourencco2020bugdoc, krishna2020cadet, artho2011iterative, zhang2014encore}.

\stitle{Optimization Algorithms.}
Bayesian optimization (BO) 
is used to optimize expensive black box functions (see \cite{frazier2018tutorial,shahriari2015taking,brochu2010tutorial,lizotte2008practical} for overviews). BO consists of a surrogate model to estimate the expensive, derivative-free objective function, and an acquisition function to determine the next best point. The most common surrogate models are Gaussian processes~\cite{schonlau1998global} (GP) and tree-structured Parzen estimators~\cite{bergstra2011algorithms,bergstra2013making} (TPE). We selected TPE since it scales linearly in the size of the set of evaluated points, whereas a GP scales cubically~\cite{bergstra2011algorithms}. Other surrogate models include random forests~\cite{hutter2011sequential} and neural networks~\cite{snoek2015scalable}.
Expected improvement~\cite{schonlau1998global} is the most common acquisition function. 

Hyperband~\cite{li2017hyperband} is a bandit-based approach for hyperparameter tuning that uses adaptive resource allocation and early-stopping to speed up random search. We did not choose Hyperband as the optimization approach since a time budget needs to be specified before running the algorithm (whereas TPE can run progressively), and our categorical variable optimizations in Section~\ref{sec:cat_enc} are designed for a sequential optimization algorithm, which  Hyperband is not.

\stitle{Categorical Bayesian Optimization.} Categorical variables in BO are often handled by one-hot encoded \cite{golovin2017google, garrido2020dealing, feurer2019hyperparameter}. However, this approach does not scale well to variables with many distinct values~\cite{ru2020bayesian}.
BO may use tree-based surrogate models (e.g., random forests~\cite{hutter2011sequential}, TPE~\cite{bergstra2011algorithms}) to handle categorical variables, however their predictive accuracy is empirically poor~\cite{garrido2020dealing,nguyen2020bayesian}.
 Other work optimizes a combinatorial search space \cite{baptista2018bayesian, deshwal2020scalable, oh2019combinatorial}, and categorical/category-specific continuous variables \cite{nguyen2020bayesian}. These works only consider categorical variables or focus on categorical variables with few distinct values, which is unsuitable for query explanation.

\vspace{-1em}

\section{Conclusion}

In this paper, we proposed \boexplain, a novel framework for explaining \sqlml queries using BO. This framework treats the \sqlml query along with an ML pipeline as a black-box which enables explanations to be derived from complex pipelines with UDFs. We considered predicates as explanations, and treated the predicate constraints as parameters to be tuned. TPE was used to tune the parameters, and we proposed a novel individual contribution encoding and warm start heuristic to improve the performance of categorical variables. We performed experiments showing that a) \boexplain can even outperform Scorpion and Macrobase for explaining SQL-only queries in certain situations, b) the proposed IC and warm start techniques were effective, c) \boexplain significantly  outperformed random search for explaining inference queries, and d) \boexplain generated useful explanations for real corrupted datasets.



\bibliographystyle{ACM-Reference-Format}
\bibliography{bibfile}

\clearpage

\appendix

{\noindent \LARGE \bf APPENDIX}

\section{Tree-structured Parzen Estimator}\label{sec:appendix_tpe}



The tree-structured Parzen estimator \cite{bergstra2011algorithms, bergstra2013making} (TPE) is a sequential model-based optimization algorithm that uses Gaussian mixture models to approximate a black-box function $f$ and the Expected Improvement \cite{schonlau1998global} acquisition function to select the next sample. At a high level, TPE splits the evaluated points into two sets: good points and bad points (as determined by the objective function). It then creates two distributions, one for each set, and finds the next point to evaluate which has a high probability in the distribution over the good points and low probability in the distribution over the bad points. We next formally define the algorithm.

Initially, $n_{\text{init}}$ samples are selected uniformally at random from the search space $\mathcal{X}$, and subsequently a model is used to guide the selection to the optimal location. TPE models each dimension of the search space independently using univariate Parzen window density estimation (or kernel density estimation) \cite{silverman1986density}. Assume for now that the search space is one-dimensional, i.e., $\mathcal{X} = [a, b]\subset \mathbb{R}$. Rather than model the posterior probability $p(y\mid x)$ directly, TPE exploits Bayes' rule, $p(y\mid x) \propto p(x\mid y)p(y)$, and models the likelihood $p(x\mid y)$ and the prior $p(y)$. To model the likelihood $p(x\mid y)$, the observations $D_t = \{(x_i , y_i = f(x_i))\}_{i=1}^t$ are first split into two sets, $D_t^g$ and $D_t^b$, based on their quality under $f$: $D_t^g$ contains the $\gamma-$quantile highest quality points, and $D_t^b$ contains the remaining points. Next, density functions $g(x)$ and $b(x)$ are created from the samples in $D_t^g$ and $D_t^b$ respectively. For each point $x\in D_t^g$, a Gaussian distribution is fit with mean $x$ and standard deviation set to the greater of the distances to its left and right neighbor.
 $g(x)$ is a uniform mixture of these distributions. The same process is performed to create the distribution $b(x)$ from the points in $D_t^b$. 
Formally, for a minimization problem, we have the likelihood $$p(x\mid y) = \begin{cases} 
g(x) & \text{ if } y < y^* \\ b(x) & \text{ if } y \ge y^*
\end{cases}$$ where $y^*$ is the $\gamma-$quantile of the observed values The prior probability is $p(y<y^*)=\gamma$.

TPE uses the prior and likelihood models to derive the Expected Improvement \cite{schonlau1998global} (EI) acquisition function. As the name suggests, EI involves computing how much improvement the objective function is expected to achieve over some threshold $y^*$ by sampling a given point. Formally, EI under some model $M$ of $f$ is defined as 
\begin{equation}\label{ei}
    EI_{y^*}(x)=\int_{-\infty}^{\infty}\max\{y^*-y, 0\}p_M(y\mid x)dy.
\end{equation}
For TPE, it follows from Equation~\ref{ei} that
\begin{equation}
    EI_{y^*}(x)\propto \left(\gamma+\frac{b(x)}{g(x)}(1-\gamma)\right)^{-1}
\end{equation}
the proof of which can be found in \cite{bergstra2011algorithms}. This means that a point with high probability in $g(x)$ and low probability in $b(x)$ will maximize the EI. To find the next point to evaluate, TPE samples $n_{EI}$ candidate points from $g(x)$. Each of these points is evaluated by $g(x)/b(x)$, and the point with the highest value is suggested as the next point to be evaluated by $f$. 

For a $d-$dimensional search space, $d > 1$, TPE is performed independently for each dimension on each iteration. The full TPE algorithm is given in Algorithm~\ref{algo:tpe}.

\begin{algorithm}[t]
\caption{Tree-structured Parzen Estimator}
\label{algo:tpe}
\DontPrintSemicolon
\SetKwInOut{Initialize}{Initialize}
\SetAlgoLined
\KwIn{$f, \mathcal{X}, n_{\text{init}}, n_{\text{iter}}, n_{EI}, \gamma$}
\KwOut{The best performing point found by TPE}

\textbf{Initialize:} Select $n_{\text{init}}$ points uniformally at random from $\mathcal{X}$, and create $D_{n_{\text{init}}} = \{(\textbf{x}_i , f(\textbf{x}_i))\}_{i=1}^{n_{\text{init}}}$\;
\For{$t\leftarrow n_{\text{init}}$ \KwTo $n_{\text{iter}}$}{
    Determine the $\gamma$-quantile point, $y^*$\;
    Split $D_t$ into $D_t^g$ and $D_t^b$ based on $y^*$\;
    \For{$i\leftarrow 1$ \KwTo $d$}{
        Estimate $g(x)$ on the i$th$ dimension of $D_t^g$\;
        Estimate $b(x)$ on the i$th$ dimension of $D_t^b$\;
        Sample $n_{EI}$ points from $g(x)$\;
        Find the sampled point $x_{t + 1}^i$ with highest $g(x)/b(x)$\; 
    }
    Update $D_{t+1} \leftarrow D_t\cup\{(\textbf{x}_{t + 1}, f(\textbf{x}_{t + 1}))\}$
}
 \KwRet{$(\textbf{x}, y)\in D_{n_{\text{iter}}}$ with the best objective function value} \;
\end{algorithm}

\section{\boexplain with Skewed Data}\label{sec:appendix_skew}

\begin{figure}
    \centering
    \includegraphics[width=0.48\textwidth]{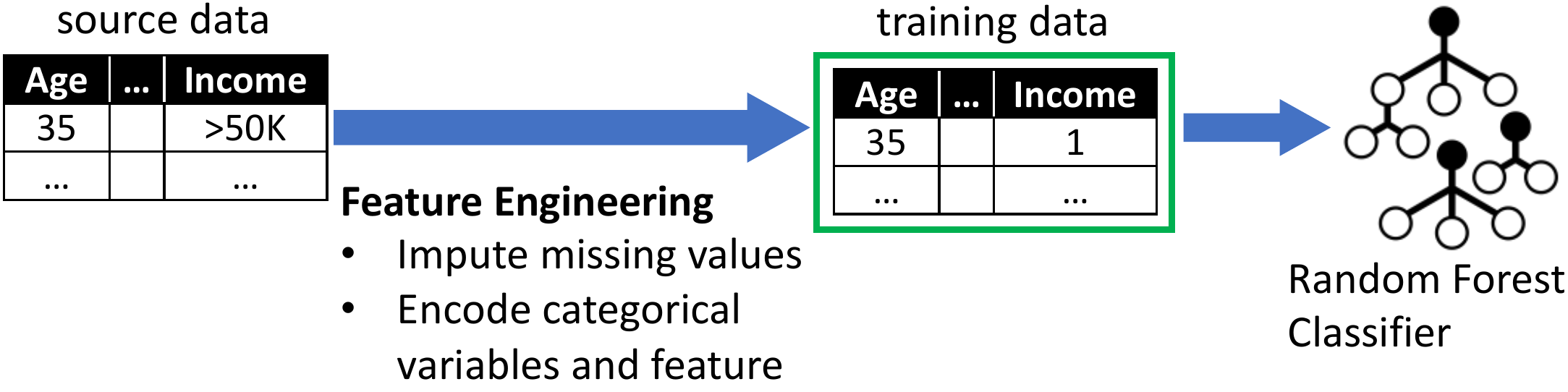}
    \caption{ML Pipeline for the Adult dataset. In this experiment, an explanation is derived from the training data.}
    \label{fig:pipeline_adult}
\end{figure}

In this section, we evaluate \boexplain in the presence of skewed data using the Adult Census Income dataset~\cite{Dua:2019}. This dataset contains 32,561 rows and 15 human variables from a 1994 census, and we use a random forest classifier to predict whether a person makes over \$50K a year. The ML pipeline is visualized in Figure~\ref{fig:pipeline_adult}, where the green box indicates that an explanation is derived from the training data. We split the data into 80\% for training and 20\% for inference. To define the ground truth corrupted data, we flipped the labels of training data tuples that satisfy the predicate $8\le\texttt{Education-Num}\le10 \wedge 30\le\texttt{Age}\le40$, which affects 16\% of the training data. On the inference data, we query the average predicted value for the group Male. To assess whether \boexplain can accurately remove the corrupted data, we define the objective function to minimize the distance between the query result on the passed data and the query result if executed on the data after filtering out the corrupted tuples. We use the two numerical search variables Education-Num and Age which have domains $[1, 16]$ and $[17, 90]$, respectively, to derive an explanation.

\begin{figure*}[!htbp]
    \centering
    \includegraphics[width=\textwidth]{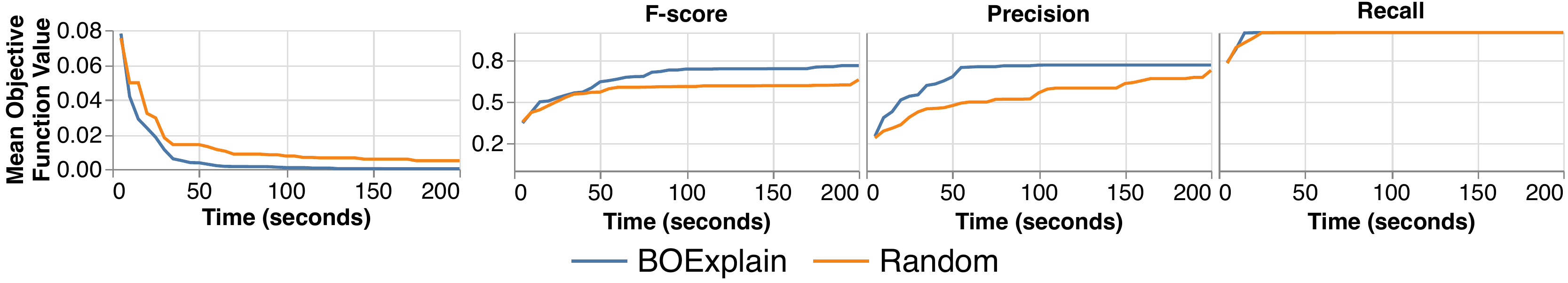}
    \caption{Adult: best objective function value, F-score, precision, and recall found at each 5 second increment, averaged over 10 runs. The goal is to minimize the objective function.}
    \label{fig:adult_result}
\end{figure*}

\begin{figure}
\centering
\begin{minipage}[t]{.22\textwidth}
  \centering
  \includegraphics[width=.99\linewidth]{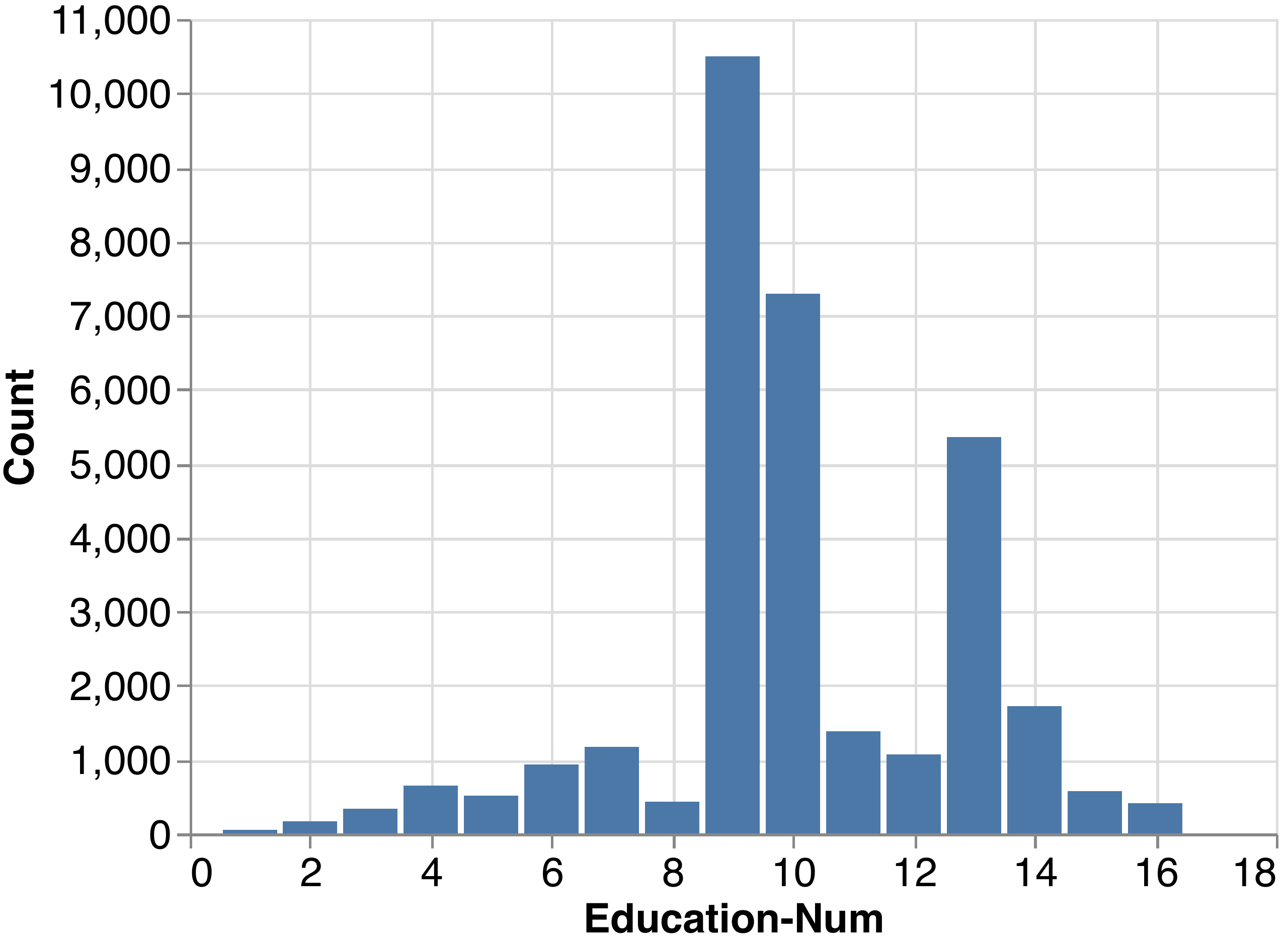}
  \captionof{figure}{Distribution of Education-Num}
  \label{fig:edu_num_dist}
\end{minipage}
\quad
\begin{minipage}[t]{.22\textwidth}
  \centering
  \includegraphics[width=.99\linewidth]{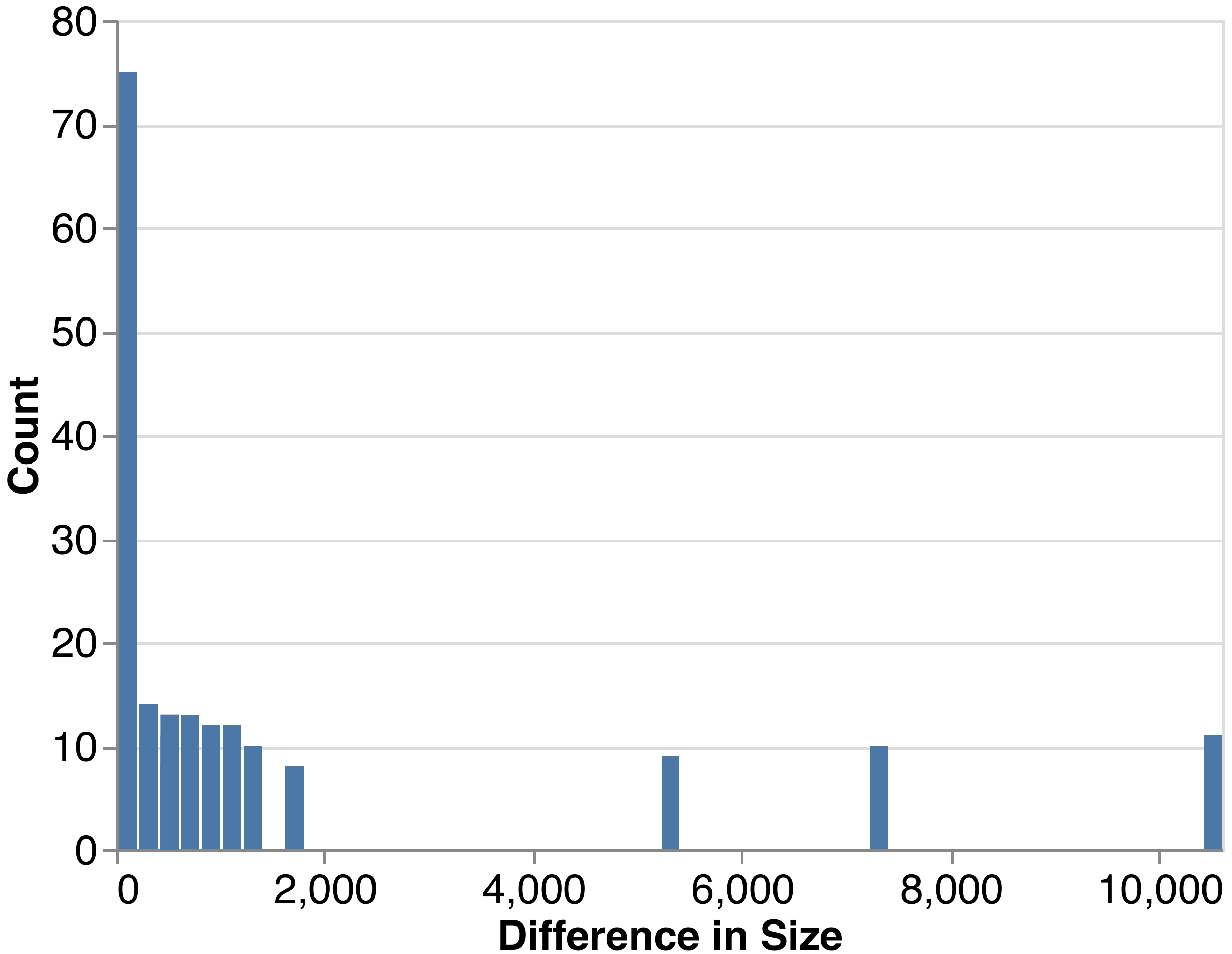}
  \captionof{figure}{Distribution of the difference in the size of sets of tuples defined by similar predicates}
  \label{fig:diff_dist}
\end{minipage}
\end{figure}

The distribution of the search variable Education-Num is given in Figure~\ref{fig:edu_num_dist}, and as we can see, some values occur much more frequently than others. Next, we will investigate whether \textit{similar} predicates define sets of tuples with similar size. Define a predicate $[w, x]$ to be similar to a predicate $[y, z]$ if $w=y$ and $x=z+1$, or $w=y+1$ and $x=z$. For each pair of similar predicates, we compute the absolute value of the difference between the number of tuples that satisfy each predicate. The distribution is plotted in Figure~\ref{fig:diff_dist}. As there is a large representation close to zero, similar predicates usually define a similar set of tuples. However, sometimes this is not the case. For example, predicate $7 \le \text{Education-Num} \le 9$ satisfies 10,501 more tuples than predicate $7 \le \text{Education-Num} \le 8$.

We ran \boexplain and random search for 200 seconds, and the results are shown in Figure~\ref{fig:adult_result}. \boexplain on average achieves an objective function result lower than Random before 45 seconds compared to Random's result at 200 seconds.
\boexplain also outperforms Random in terms of F-score and precision. 
These results show that \boexplain can still perform well in the presence of skewed data.

\end{document}